\documentclass[a4paper,11pt]{article}
\usepackage{jheppub} 
\usepackage{epsfig}
\usepackage{graphicx}
\usepackage[T1]{fontenc}
\def\be{\begin{equation}}
\def\ee{\end{equation}}
\def\bea{\begin{array}}
\def\eea{\end{array}}
\def\beqa{\begin{eqnarray}}
\def\eeqa{\end{eqnarray}}
\def\beqas{\begin{eqnarray*}}
\def\eeqas{\end{eqnarray*}}

\def\bp{\begin{picture}}
\def\ep{\end{picture}}
\def\bc{\begin{center}}
\def\ec{\end{center}}
\def\bfig{\begin{figure}}
\def\efig{\end{figure}}

\def\bit{\begin{itemize}}
\def\eit{\end{itemize}}
\def\nn{\nonumber}
\def\f{\frac}

\def\[{\left[}
\def\]{\right]}
\def\({\left(}
\def\){\right)}

\def\..{\left.}
\def\.{\right.}
\def\tl{\tilde}
\def\ra{\rightarrow}

\def\tm{\times}

\def\al{\alpha}
\def\bt{\beta}

\def\ep{\epsilon}

\def\ga{\gamma}
\def\pa{\partial}
\def\pr{\prime}

\title{\boldmath Traveling Wave Form Description for Dirac Field and Its Deduction To Pauli Equation Type Forms in Quantum Mechanics}
\author[a]{Fei Wang\note{Corresponding author.}}
\affiliation[a]{Department of Physics, Zhengzhou University, No. 100 Science Avenue, ZhengZhou 450001, P. R. China}
\emailAdd{feiwang@zzu.edu.cn}

\abstract{  We derive an equivalent traveling wave form description for Dirac field. In the non-relativistic limit, such form can reduce to inverse-Galilean transformed Schrodinger-type equation. We find that, the resulting two-component Schrodinger-type equation from the reduction of traveling wave form description of Dirac field is different to the naive Galilean transformed Schrodinger equation. Taking into account the interactions of the system to electromagnetic field by adding proper forms of covariant derivative, the traveling wave form description for Pauli equation can be similarly obtained in the non-relativistic limit. Such descriptions allow one to choose arbitrary convenient reference frame for quantum system involving spins. Using Bargmann-Wigner formalism for field with arbitrary spin $s\geq 1/2$, which satisfy Dirac-type equations in all its indices, the traveling wave description for such a field can be similarly obtained from the traveling wave form description of Dirac field, for example, for the spin-3/2 Rarita-Schwinger field and spin-2 gravitational field.
}
\begin{document}
\maketitle
\flushbottom

\section{Introduction}
In the ordinary description of electromagnetic field, the electric field and magnetic field are stationary in the sense that their space variables are independent of time. It is still controversial whether the traveling wave description of electromagnetic fields~\cite{rozov,wzl,wzl2} with simple inverse Galilean coordinate variables~\cite{wx} can be consistent. Recently, an alternative equivalent (traveling wave type) description of Maxwell's equations is derived with special relativity in~\cite{WY} and ~\cite{tjl,qw,wzl3}, which can reduce to the (extended) Hertz-form equations~\cite{hertz,pauli,rozov} from non-relativistic (low speed) expansion. This means that the traveling wave type description for electromagnetic field satisfies the extended-Hertz type Maxwell's equations instead of ordinary Maxwell's equations.

 In the language of quantum field theory, the electromagnetic field corresponds to massless spin-1 photon. It is well known that the  spin-0 and spin-1/2 fields are described by the relativistic Klein-Gordon equation and Dirac equation, respectively. So, it is interesting to survey if the previous formalism for traveling wave type description of spin-1 field can be generalized to other fields. It is easy to see that the Klein-Gordon equation will not be changed as the scalar fields transformed in a rather trivial way under Lorentz transformation. The spin-1/2 Dirac field, however, transforms non-trivially under Lorentz transformation and can possibly have an alternative description. So, we would like to survey the traveling wave description of Dirac field and discuss its possible consequences.

The Dirac equation is a relativistic field equations that describe the generation and annihilation of particles. It can act as a single-particle quantum mechanical wave equation in its original incarnation, although bothered by inconsistencies such as the Klein paradox. The single-particle wave function of Dirac equation can reduce to ordinary Schrodinger wave function in the non-relativistic limit. So, the traveling wave description of Dirac field can reduce to traveling wave description of Schrodinger equation, which amounts to the Galilean transformation of Schrodinger equation when the traveling wave speed is much smaller than the light speed.

It is well known that the Schrodinger equation is of non-relativistic nature, which is applicable
for microscopic particles with speed much smaller than the light speed. However, the Schrodinger is not covariant under the Galilean boosts unless we redefine the Galilean transformation by extending it with a phase shift~\cite{schrodinger:galilean,schrodinger:galilean2}. So, it is interesting to see if new perspective can be given on the Galilean type transformations for Schrodinger equations, especially for a two-component system that is observed in a new moving reference frame. Schrodinger equation can be regarded as the non-relativistic limit of Klein-Gordon and Dirac equations. As ordinary Dirac equation can reduce to the Schrodinger equation in the non-relativistic limit, it is therefore interesting to see what form the traveling wave description of the spin-1/2 field can reduce to in the non-relativistic limit with small traveling wave speed.

This paper is organized as follows. In Sec.~\ref{sec-2}, we derive the traveling wave type description for Dirac field and its Schrodinger equation type form in the non-relativistic limit. In Sec.~\ref{sec-2}, we include the interaction of the system to electromagnetic field by adopting proper forms of covariant derivative in the traveling wave type description of Dirac equation. The resulting traveling wave type description for Pauli equation (in the non-relativistic limit) is derived.
Sec.~\ref{conclusion} contains our conclusions.

\section{\label{sec-2} Traveling wave type description for Dirac field and its non-relativistic limit}
The Dirac equation can be written as
\beqa
(i \ga^\mu \pa_\mu -m)\psi(x)=0~,
\eeqa
with
\beqa
\ga^\mu=\(\bea{cc}0& \sigma^\mu\\\bar{\sigma}^\mu &0 \eea\).
\eeqa
Here $\sigma^\mu\equiv(1,\sigma^i)$ and $\bar{\sigma}^\mu=(1,-\sigma^i)$.
It is well known that Dirac equation is invariant under Lorentz transformations. The transformation of $\psi(x)$ under Lorentz transformation
\beqa
x^\pr_\mu=\Lambda_\mu^\nu x_\nu~,~~~~~~\left\{
\bea{c}\vec{x}^\pr=\vec{x}+\vec{v} t~,\\ t^\pr=t+ \vec{v}\cdot \vec{c} ~,\eea\right.
\eeqa
is given by
\beqa
\psi^\pr(x^\pr)=\Lambda_{1/2} \psi(x)~.
\eeqa
In the discussion, we use the natural unit to set $\hbar=c=1$ and will change back in the end.
Therefore, we have
\beqa
&&(i \ga^\mu \pa^\pr_\mu -m)\psi^\pr(x^\pr)=0~,\nn\\
&\Rightarrow& (i \ga^\mu \pa^\pr_\mu -m)\Lambda_{1/2} \psi (x)=0~,\nn\\
&\Rightarrow& (i \ga^\mu \pa^\pr_\mu -m)\Lambda_{1/2} \tl{\psi} (x^\pr)=0~,\nn\\
&\Rightarrow& (i \ga^\mu \pa_\mu -m)\Lambda_{1/2} \tl{\psi} (x)=0~,
\label{dirac:tw}
\eeqa
after defining
\beqa
\tl{\psi}(x^\pr)\equiv {\psi}[x(x^\pr)]=\psi(\Lambda^{-1}x^\pr)~,
\eeqa
with the derivative $\pa^\pr_\mu$ for $x^\pr$ coordinates. This is the equivalent traveling-wave type description for spinor, as the non-relativistic limit of $\tl{\psi}(x^\pr)$
\beqa
\tl{\psi}(\vec{x}^\pr,t^\pr)=\psi(\Lambda^{-1}x^\pr)\approx \psi(\vec{x}^\pr-\vec{v} t^\pr,t^\pr) ~,
\eeqa
is just the inverse Galilean transformed Dirac field.

The expression of $\Lambda_{1/2}$ is given by
\beqa
\Lambda_{1/2}=\exp\(-\f{i}{2}\omega_{\mu\nu}S^{\mu\nu}\)~,~~~{\rm with}~ S^{\mu\nu}=\f{i}{4}\[\ga^\mu,\ga^\nu\]~,
\eeqa
and the transformation parameter $\omega_{\mu\nu}$ is given from the Lorentz transformation of coordinates $x^\pr=\Lambda x$ with
\beqa
\Lambda=\exp\(-\f{i}{2}\omega_{\mu\nu}\mathcal{J}^{\mu\nu}\)~,~~
~~{\rm with}~\(\mathcal{J}^{\mu\nu}\)_{\al\bt}=i(\delta_{\al}^\mu\delta_{\bt}^\nu-\delta^\mu_\bt\delta^\nu_\alpha).
\eeqa
The transformation $\Lambda$ can be rewritten as
\beqa
\Lambda=\(\bea{cc}\cosh \eta~ & ~\sinh \eta \\\sinh \eta~&~ \cosh \eta  \eea\).
\eeqa
for boost in the $x^3$ direction and the corresponding $\Lambda_{1/2}$ is given as
\beqa
\Lambda_{1/2}&=&\exp\[-\f{1}{2}\eta\(\bea{cc}\sigma^3& 0\\0&-\sigma^3\eea\)\]~,\nn\\
&=&\cosh\(\f{\eta}{2}\)\(\bea{cc}~1~&~0~\\~0~&~1~\eea\)-\sinh\(\f{\eta}{2}\)\(\bea{cc}\sigma^3&0\\0&-\sigma^3\eea\)~,\nn\\
&=&\(\bea{cc}e^{\f{\eta}{2}}\(\f{1-\sigma_3}{2}\)
+e^{-\f{\eta}{2}}\(\f{1+\sigma_3}{2}\)~ & ~0  \\
0~&~e^{\f{\eta}{2}}\(\f{1+\sigma_3}{2}\)
+e^{-\f{\eta}{2}}\(\f{1-\sigma_3}{2}\) \eea\).\eeqa
In the small speed approximation, we have
\beqa
\Lambda\approx \(\bea{cc}1~ & ~ \eta \\ \eta~&~ 1\eea\)~~~~~~~{\rm with}~~\eta=\tanh^{-1}\(\f{v}{c}\)\approx \f{v}{c}=v~,
\eeqa
and
\beqa
\Lambda_{1/2}&\approx&\(\bea{cc}~1_{2\tm 2}~&~0~\\~0~&~1_{2\tm 2}~\eea\)-\(\f{\eta}{2}\)\(\bea{cc}\sigma^3&0\\0&-\sigma^3\eea\)~.
\eeqa
for the unit with $c=1$.
So we have
\beqa
\Lambda_{1/2}\tl{\psi}(x)&=&\(\bea{cc}~1_{2\tm 2}-\(\f{\eta}{2}\)\sigma^3~&~0~\\~0~&~
 1_{2\tm 2}+\(\f{\eta}{2}\)\sigma^3~\eea\)\(\bea{c}\tl{\psi}_L(x)\\\tl{\psi}_R(x)\eea\)\nn\\
 &=&\(\bea{c}\tl{\psi}_L(x)-\(\f{\eta}{2}\)\sigma^3\tl{\psi}_L(x)
 \\\tl{\psi}_R(x)+\(\f{\eta}{2}\)\sigma^3\tl{\psi}_R(x)
 \eea\).
\eeqa

In two-component conventions, we can rewrite the Dirac spinor as
\beqa
\psi(x)=\(\bea{c}\psi_L(x)\\ \psi_R(x)\eea\)~,~~~\tl{\psi}(x)=\(\bea{c}\tl{\psi}_L(x)\\\tl{\psi}_R(x)\eea\)~.
\eeqa
The Dirac equation with function form $\tl{\psi}$
\beqa
(i \ga^\mu \pa_\mu -m)\Lambda_{1/2} \tl{\psi} (x)=0~,
\eeqa
can be rewritten in two-components
\beqa
i\(\pa_0-\vec{\sigma}\cdot{\nabla}\)\[\tl{\psi}_L(x)-\(\f{\eta}{2}\)\sigma^3\tl{\psi}_L(x)\]
&=&m\[\tl{\psi}_R(x)+\(\f{\eta}{2}\)\sigma^3\tl{\psi}_R(x)\]~,\nn\\
i\(\pa_0+\vec{\sigma}\cdot{\nabla}\)\[\tl{\psi}_R(x)+\(\f{\eta}{2}\)\sigma^3\tl{\psi}_R(x)\]
&=&m\[\tl{\psi}_L(x)-\(\f{\eta}{2}\)\sigma^3\tl{\psi}_L(x)\]~.
\label{Dirac:Travel}
\eeqa
For massless spinor, we have
\beqa
i\(\pa_0-\vec{\sigma}\cdot{\nabla}\)\[\tl{\psi}_L(x)-\(\f{\eta}{2}\)\sigma^3\tl{\psi}_L(x)\]&=&0~,\nn\\
i\(\pa_0+\vec{\sigma}\cdot{\nabla}\)\[\tl{\psi}_R(x)+\(\f{\eta}{2}\)\sigma^3\tl{\psi}_R(x)\]&=&0~.
\eeqa
The general boost case can be obtained by the replacement
\beqa
v\sigma^3\ra \(\vec{\sigma}\cdot\vec{v}\).
\eeqa
So, we obtain the massless spinor case
\beqa
i\[\pa_0-\(\vec{\sigma}\cdot{\nabla}\)-\f{1}{2}\(\vec{v}\cdot\vec{\sigma}\)\pa_0
+\f{1}{2}\(\vec{\sigma}\cdot{\nabla}\)\(\vec{v}\cdot\vec{\sigma}\)\]\tl{\psi}_L(x)&=&0~,\nn\\
i\[\pa_0+\(\vec{\sigma}\cdot{\nabla}\)+\f{1}{2}\(\vec{v}\cdot\vec{\sigma}\)\pa_0
+\f{1}{2}\(\vec{\sigma}\cdot{\nabla}\)\(\vec{v}\cdot\vec{\sigma}\)\]\tl{\psi}_R(x)&=&0~,
\eeqa
Using the formula
\beqa
\({\vec{P}}\cdot\vec{\sigma}\)\(\vec{v}\cdot\vec{\sigma}\)
=\(\vec{P}\cdot{\vec{v}}\)+i\(\vec{P}\tm {\vec{v}}\)\cdot\vec{\sigma}~,
\eeqa
we will arrive at
\beqa
i\left\{\pa_0-\(\vec{\sigma}\cdot{\nabla}\)-\f{1}{2}\(\vec{\sigma}\cdot\vec{v}\)\pa_0
+\f{1}{2}\[(\vec{v}\cdot\nabla)-i\vec{\sigma}\cdot \(\vec{v}\tm \nabla\) \]\right\}\tl{\psi}_L(x)&=&0~,\nn\\
i\left\{\pa_0+\(\vec{\sigma}\cdot{\nabla}\)+\f{1}{2}\(\vec{\sigma}\cdot\vec{v}\)\pa_0
+\f{1}{2}\[(\vec{v}\cdot\nabla)-i\vec{\sigma}\cdot \(\vec{v}\tm \nabla\) \]\right\}\tl{\psi}_R(x)&=&0~,
\label{massless:NR}
\eeqa
for the traveling wave type massless two-component Weyl equations.
We know that the naive Galilean transformation adopt the replacements
\beqa
\pa_0\ra \pa_0+\vec{v}\cdot{\nabla}~,~~~\nabla\ra \nabla+{\vec{v}}\pa_0~.
\label{Galilean}
\eeqa
Comparing with the Galilean transformed massless two-component Weyl equations, we will find that there are additional $1/2$ factors for the $\vec{v}\cdot{\nabla}$ term within $\pa_0$ replacement and for the ${\vec{v}}\pa_0$ term within $\nabla$ replacement in the traveling wave type Weyl equation for $\tl{\psi}_R$, respectively. An additional sign difference appear for the $\vec{v}\cdot{\nabla}$ term in the traveling wave type Weyl equation for $\tl{\psi}_L$. Besides, new additional $\vec{\sigma}\cdot(\vec{v}\cdot\nabla)$ terms also emerge.

For $m\neq 0$, traveling wave type two-component equations can be written as
\beqa
i\left\{\pa_0-\(\vec{\sigma}\cdot{\nabla}\)-\f{1}{2}\(\vec{\sigma}\cdot\vec{v}\)\pa_0
+\f{1}{2}\[(\vec{v}\cdot\nabla)-i\vec{\sigma}\cdot \(\vec{v}\tm \nabla\) \]\right\}\tl{\psi}_L(x)&=&m\[\tl{\psi}_R(x)+\f{1}{2}\(\vec{\sigma}\cdot\vec{v}\)\tl{\psi}_R(x)\]~,\nn\\
i\left\{\pa_0+\(\vec{\sigma}\cdot{\nabla}\)+\f{1}{2}\(\vec{\sigma}\cdot\vec{v}\)\pa_0
+\f{1}{2}\[(\vec{v}\cdot\nabla)-i\vec{\sigma}\cdot \(\vec{v}\tm \nabla\) \]\right\}\tl{\psi}_R(x)&=&m\[\tl{\psi}_L(x)-\f{1}{2}\(\vec{\sigma}\cdot\vec{v}\)\tl{\psi}_L(x)\]~.\nn\\
\eeqa

Assuming that the relevant energy and momentum are much smaller than the rest mass of the particle $m$, we can obtain the non-relativistic limit of such traveling wave type Dirac equations by picking out the rest mass contribution
\beqa
\tl{\psi}(x)&=&\(\bea{c}\tl{\psi}_L(x)\\\tl{\psi}_R(x)\eea\)=\(\bea{c}\tl{\Psi}_L(x)\\\tl{\Psi}_R(x)\eea\)e^{-im t},
\eeqa
with $\tl{\Psi}_L(x),\tl{\Psi}_R(x)$ are slow varying fields.
So, the traveling-wave description Dirac equations eq.(\ref{Dirac:Travel}) can be simplified
\beqa
i\(\pa_0-\vec{\sigma}\cdot{\nabla}\)\[\tl{\Psi}_L(x)-\(\f{\eta}{2}\)\sigma^3\tl{\Psi}_L(x)\]
&=&m\[\tl{\Psi}_R(x)+\(\f{\eta}{2}\)\sigma^3\tl{\Psi}_R(x)-\tl{\Psi}_L(x)
+\(\f{\eta}{2}\)\sigma^3\tl{\Psi}_L(x)
\]~,\nn\\
i\(\pa_0+\vec{\sigma}\cdot{\nabla}\)\[\tl{\Psi}_R(x)+\(\f{\eta}{2}\)\sigma^3\tl{\Psi}_R(x)\]
&=&m\[\tl{\Psi}_L(x)-\(\f{\eta}{2}\)\sigma^3\tl{\Psi}_L(x)-\tl{\Psi}_R(x)
-\(\f{\eta}{2}\)\sigma^3\tl{\Psi}_R(x)
\]~,\nn\\
\eeqa
After rearranging the terms and combining both equations, we have
\beqa
i\[ \pa_0+\(\f{\eta}{2}\)\(\vec{\sigma}\cdot{\nabla}\)\sigma^3 \]\(\tl{\Psi}_L(x)+\tl{\Psi}_R(x)\)&=&
i\[\vec{\sigma}\cdot{\nabla}+\(\f{\eta}{2}\)\sigma^3\pa_0\]\(\tl{\Psi}_L(x)-\tl{\Psi}_R(x)\)~,\nn\\
\label{Travel:first}
\eeqa
and
\beqa
&&-i\[\pa_0+\(\f{\eta}{2}\)\(\vec{\sigma}\cdot{\nabla}\)\sigma^3 \]\(\tl{\Psi}_L(x)-\tl{\Psi}_R(x)\)
+i\[\vec{\sigma}\cdot{\nabla}+\(\f{\eta}{2}\)\sigma^3\pa_0\]\(\tl{\Psi}_L(x)+\tl{\Psi}_R(x)\)\nn\\
&=&2m\[\(\tl{\Psi}_L(x)-\tl{\Psi}_R(x)\)-\f{\eta}{2}\sigma^3\(\tl{\Psi}_L(x)+\tl{\Psi}_R(x)\)\]~.
\eeqa
For slow varying $\tl{\Psi}_L(x),\tl{\Psi}_R(x)$, we have
\beqa
\[1+i\(\f{\eta}{4m}\)\(\vec{\sigma}\cdot{\nabla}\)\sigma^3 \]\(\tl{\Psi}_L(x)-\tl{\Psi}_R(x)\)\approx \left\{\f{\eta}{2}\sigma^3+\f{i}{2m}\[\vec{\sigma}\cdot{\nabla}+\(\f{\eta}{2}\)\sigma^3\pa_0\]\right\}\(\tl{\Psi}_L(x)+\tl{\Psi}_R(x)\).\nn\\
\eeqa
So we have
\beqa
\(\tl{\Psi}_L(x)-\tl{\Psi}_R(x)\)\approx \left\{\f{\eta}{2}\sigma^3+i\f{\vec{\sigma}\cdot{\nabla}}{2m}+\f{i}{2m}\(\f{\eta}{2}\)\sigma^3\pa_0\right\}\(\tl{\Psi}_L(x)+\tl{\Psi}_R(x)\),
\label{small:component}
\eeqa
after neglecting the $v^2$ and $1/m^2$ terms. So "$\tl{\Psi}_L(x)-\tl{\Psi}_R(x)$" is the "small" component while "$\tl{\Psi}_L(x)+\tl{\Psi}_R(x)$" is the "big" component.
Substituting the expression (\ref{small:component}) into (\ref{Travel:first}), we have
\beqa
i\[\pa_0+ \(\f{\eta}{2}\)\(\vec{\sigma}\cdot{\nabla}\)\sigma^3\] \tl{\Psi}&=&-\f{\nabla^2}{2m}\tl{\Psi}+{i}\f{\eta}{2}\(\vec{\sigma}\cdot{\nabla}\)\sigma^3\tl{\Psi}
-\f{\eta}{4m}\(\sigma^3\pa_0\)\(\vec{\sigma}\cdot{\nabla}\)\tl{\Psi}
-\f{\eta}{4m}\(\vec{\sigma}\cdot{\nabla}\)\(\sigma^3\pa_0\)\tl{\Psi}~,\nn\\
\label{NR:Dirac}
\eeqa
after defining
\beqa
\(\tl{\Psi}_L(x)-\tl{\Psi}_R(x)\)=\tl{\Phi}(x),~~~\(\tl{\Psi}_L(x)+\tl{\Psi}_R(x)\)=\tl{\Psi}(x)~,
\eeqa
and neglecting the $v^2$ terms.
The case with general boost direction requires the replacement
\beqa
v\sigma^3\ra \(\vec{v}\cdot\vec{\sigma}\).
\eeqa
Thus, we have
\beqa
\eta\(\vec{\sigma}\cdot{\nabla}\)\sigma^3\ra
\(\vec{\sigma}\cdot{\nabla}\)\(\vec{\sigma}\cdot\vec{v}\)
=\[\({\vec{v}}\cdot\nabla\)-i\vec{\sigma}\cdot\(\vec{v}\tm\nabla\)\]~.
\eeqa
So, the eq.(\ref{NR:Dirac}) can be rewritten as
\beqa
i\hbar\f{\pa}{\pa t} \tl{\Psi}(x)&=&-\f{\hbar^2}{2m}{\nabla^2}\tl{\Psi}(x)-\f{\hbar^2}{2m c^2}\({\vec{v}}\cdot\nabla\)\f{\pa}{\pa t}\tl{\Psi}~,
\eeqa
after changing back from the natural unit. In the non-relativistic limit, the traveling wave type two-component wavefunction $\tl{\Psi}(x)$ in the previous equation can be approximated by
\beqa
\tl{\Psi}(x)\approx \Psi(\vec{x}-\vec{v}t,t)~.
\eeqa
with $\Psi$ corresponding to the ordinary two-component wavefunction that satisfy the Schrodinger equation.

In the Schrodinger equation, naive replacement for Galilean transformation
\beqa
\pa_0\ra \pa_0+\vec{v}\cdot{\nabla}~,~~~\nabla\ra \nabla+\f{\vec{v}}{c^2}\pa_0~,
\eeqa
will change the Schrodinger equation into
\beqa
i\hbar(\f{\pa}{\pa t}+\vec{v}\cdot{\nabla})\Psi(\vec{x}-\vec{v}t,t)&=&
-\f{\hbar^2}{2m}\(\nabla+\f{\vec{v}}{c^2}\f{\pa}{\pa t}\)^2\Psi(\vec{x}-\vec{v}t,t)~,\nn\\
&\approx&-\f{\hbar^2}{2m}{\nabla^2}\Psi(\vec{x}-\vec{v}t,t)-\f{\hbar^2}{m c^2} (\vec{v}\cdot\nabla)\f{\pa}{\pa t}\Psi(\vec{x}-\vec{v}t,t)~,
\eeqa
after neglecting the $v^2$ term.
\textbf{So we can see that the equation for traveling wave description $\Psi^\pr(\vec{x}^\pr-\vec{v}t,t)$ derived from non-relativistic limit of traveling wave type Dirac equation is differen from the expression from naive Galilean transformation}.

In the Bargmann-Wigner formulation, a field of rest mass $m$ and spin $s\geq 1/2$ can be represented by a completely symmetric multi-spinor $\Psi_{\al_1\cdots\al_{2s}}(x)$ of rank $2s$ that satisfy Dirac-type equations in all its indices
\beqa
i(\ga^\mu\pa_\mu-m)_{\al_1 \al_1^\pr} \Psi_{\al_1^\pr\cdots\al_{2s}}(x)&=&0~,\nn\\
\cdots\cdots\cdots\cdots\cdots&\cdots&\\
i(\ga^\mu\pa_\mu-m)_{\al_{2s} \al_{2s}^\pr} \Psi_{\al_1\cdots\al_{2s}^\pr}(x)&=&0~,
\eeqa
After Lorentz transformation, we can use similar deductions in eq.(\ref{dirac:tw}) to give
\beqa
i(\ga^\mu\pa_{\mu}-m)_{\al_{k} \al_{k}^\pr} \(\Lambda_{1/2}\)_{\al_k^\pr \al_l^\pr}\tl{\Psi}_{\al_1\cdots \al_l^\pr \cdot\al_{2s}}(x)=0~,
\eeqa
with
\beqa
\tl{\Psi}_{\al_1\cdots \al_l^\pr \cdot\al_{2s}}(x^\pr)={\Psi}_{\al_1\cdots \al_l^\pr \cdot\al_{2s}}[x(x^\pr)]={\Psi}_{\al_1\cdots \al_l^\pr \cdot\al_{2s}}(\Lambda^{-1} x^\pr).
\eeqa
\section{\label{sec-3}Traveling wave form description for Pauli equation}
Free Dirac equation can be extended to include the couplings of the fermion to various gauge fields by replacing the ordinary derivative with covariant derivative.
Given the fact that the color freedom of $SU(3)_c$ confines and the $W,Z$ gauge bosons can be integrated out in the low energy region, we will only concentrate on the coupling of fermion to the $U(1)_Q$ electromagnetic field.

The Dirac equation with covariant derivative involving electromagnetic field is given as
\beqa
i \ga^\mu (\pa_\mu-ie A_\mu(x))\psi(x) -m\psi(x)=0~,
\eeqa
After Lorentz transformation $x^\pr=\Lambda x$, the ordinary Dirac equation changes into
\beqa
\[i \ga^\mu \(\pa^\pr_\mu-ie A_\mu^\pr(x^\pr)\) -m\]\psi^\pr(x^\pr)&=&0~,\nn\\
\Rightarrow~~\[i \ga^\mu \(\pa^\pr_\mu-ie \Lambda A_\mu (x)\) -m\]\Lambda_{1/2}\psi(x)&=&0~,\nn\\
\Rightarrow~\[i \ga^\mu \(\pa^\pr_\mu-ie \Lambda \mathcal{A}_\mu ( x^\pr)\) -m\]\Lambda_{1/2}\tl{\psi}(x^\pr)&=&0~,\nn\\
\Rightarrow~~~~\[i \ga^\mu \(\pa_\mu-ie \Lambda \mathcal{A}_\mu ( x)\) -m\]\Lambda_{1/2}\tl{\psi}(x)&=&0~,
\eeqa
with
\beqa
{A}_\mu(x)={A}_\mu[x(x^\pr)]\equiv \mathcal{A}_\mu(x^\pr).
\eeqa

From the four-vector form of $A_\mu(x)=(\phi[x(x^\pr)],\vec{A}[x(x^\pr)])
\equiv(\phi^\pr(x^\pr),\vec{\mathcal{A}}(x^\pr)) $, we can have
\beqa
\Lambda \mathcal{A}_\mu (x)\equiv\Lambda ({\phi}^\pr,\vec{\mathcal{A}})
=\(\phi^\pr+{\vec{v}\cdot\vec{\mathcal{A}}},\vec{\mathcal{A}}+{\vec{v}}\phi^\pr\)~.
\eeqa
The resulting covariant derivative can be obtained by the replacement
\beqa
\nabla \ra \nabla-i e \(\vec{\mathcal{A}}+{\vec{v}}\phi^\pr\),
\eeqa
and replacement
\beqa
\pa_0\ra \pa_0-ie \(\phi^\pr+{\vec{v}\cdot\vec{\mathcal{A}}}\).
\eeqa

Using the Lorentz transformation for electromagnetic fields and the non-relativistic low speed expansion up to ${\cal O}(v)$ terms, we have
\beqa
\vec{B}^\pr(x^\pr)&=&\nabla^\pr \tm \vec{A}^\pr(x^\pr)~,\nn\\
\Rightarrow(\vec{B}(x)+{{\vec{v}}}\tm\vec{E}(x))&\approx&\nabla^\pr \tm \[\vec{A}(x)+{\vec{v}}\phi(x)\]~,~~~~~~~~~~~\nn\\
\Rightarrow(\vec{\mathcal{B}}(x^\pr)+{{\vec{v}}}\tm\vec{\mathcal{E}}(x^\pr))&=&\nabla^\pr \tm \[\vec{\mathcal{A}}(x^\pr)+{\vec{v}}\phi^\pr(x^\pr)\]~,~~~~~~~~
\eeqa
and
\beqa
\vec{E}^\pr(x^\pr)&=&-\nabla^\pr \phi^\pr(x^\pr)-\f{\pa }{\pa t^\pr}\vec{A}^\pr(x^\pr)~,\nn\\
\Rightarrow (\vec{E}(x)-{\vec{v}}\tm\vec{B}(x))&\approx&
-\nabla^\pr\[\phi(x)+{\vec{v}\cdot\vec{A}(x)}\]-\f{\pa }{\pa t^\pr}\[\vec{A}(x)+{\vec{v}}\phi(x)\]~,\nn\\
\Rightarrow(\vec{\mathcal{E}}(x^\pr)-{\vec{v}}\tm\vec{\mathcal{B}}(x^\pr))
&=&-\nabla^\pr\[\phi^\pr(x^\pr)+{\vec{v}\cdot\vec{\mathcal{A}}(x^\pr)}\]
-\f{\pa }{\pa t^\pr}\[\vec{\mathcal{A}}(x^\pr)+{\vec{v}}\phi^\pr(x^\pr)\]~.
\eeqa
So, we have
\beqa
\nabla \tm \(\vec{\mathcal{A}}+{\vec{v}}\phi^\pr\)=\vec{\mathcal{B}}+{{\vec{v}}}\tm\vec{\mathcal{E}}~,~~
-\nabla\(\phi^\pr+\vec{v}\cdot\vec{\mathcal{A}}\)
=\vec{\mathcal{E}}-{\vec{v}}\tm\vec{\mathcal{B}}+\f{\pa}{\pa t}\[\vec{\mathcal{A}}+{\vec{v}}\phi^\pr\]~.
\label{Amuexpressions}
\eeqa
which, after returning back from natural unit, can be written as
\beqa
\nabla \tm \(\vec{\mathcal{A}}+\f{\vec{v}}{c^2}\phi^\pr\)&=&\vec{\mathcal{B}}+\f{{\vec{v}}}{c^2}\tm\vec{\mathcal{E}}~,\nn\\
~
-\nabla\(\phi^\pr+\vec{v}\cdot\vec{\mathcal{A}}\)
&=&\vec{\mathcal{E}}-{\vec{v}}\tm\vec{\mathcal{B}}+\f{1}{c}\f{\pa}{\pa t}\[\vec{\mathcal{A}}+\f{\vec{v}}{c^2}\phi^\pr\].
\eeqa

New expressions for massless spinor coupling to electromagnetic fields can be obtained by the covariant derivative replacement with $\vec{\mathcal{A}},\phi^\pr$ in eq.(\ref{massless:NR}).

We would like to derive the traveling wave description for Pauli equation
\beqa
i\hbar\f{\pa}{\pa t}\Psi=\[\f{1}{2m}(\vec{P}+e\vec{A})^2+\f{e\hbar}{2mc}\vec{\sigma}\cdot \vec{B}-e\phi\]\Psi~,
\eeqa
which is the non-relativistic limit for Dirac equation coupling to electromagnetic fields. We follow the deductions in previous section except that the covariant derivatives and $e\phi^\pr\ll m$ are used.
The eq.(\ref{NR:Dirac}), after replacing the covariant derivative, will be changed into
\beqa
&&\[i\pa_0-e \(\phi^\pr+{\vec{v}\cdot\vec{\mathcal{A}}}\)\] \tl{\Psi}\nn\\
&=&
-\f{1}{2m}\[\vec{\sigma}\cdot\(\nabla-ie (\vec{\mathcal{A}}+{\vec{v}}\phi^\pr)\)\]^2\tl{\Psi}-\f{1}{4m}\[\vec{\sigma}\cdot\(\nabla-ie (\vec{\mathcal{A}}+{\vec{v}}\phi^\pr)\)\]\(\vec{\sigma}\cdot \vec{v}\)
\f{\pa}{\pa t}\tl{\Psi}\nn\\
&-&\f{1}{4m}\(\vec{\sigma}\cdot \vec{v}\)\f{\pa}{\pa t}\[\vec{\sigma}\cdot\(\nabla-ie (\vec{\mathcal{A}}+{\vec{v}}\phi^\pr)\)\]\tl{\Psi}\nn\\
&=&-\f{1}{2m}\[\nabla-ie (\vec{\mathcal{A}}+{\vec{v}}\phi^\pr)\]^2\tl{\Psi}
-\f{1}{2m}e\vec{\sigma}\cdot\[\vec{\mathcal{B}}+{{\vec{v}}}\tm\vec{\mathcal{E}}\]\tl{\Psi}
-\f{1}{2m}\({\vec{v}}\cdot\[\nabla-ie (\vec{\mathcal{A}}+{\vec{v}}\phi^\pr)\]\)\f{\pa}{\pa t}\tl{\Psi}~\nn\\
&+&\f{ie}{4m}\left\{\(\vec{v}\cdot \f{\pa}{\pa t}(\vec{\mathcal{A}}+{\vec{v}}\phi^\pr)\)+i\vec{\sigma}\cdot \[\vec{v}\tm\f{\pa}{\pa t}(\vec{\mathcal{A}}+{\vec{v}}\phi^\pr)\]\right\}\tl{\Psi}~.
\eeqa
Using the expressions in eq(\ref{Amuexpressions}) and neglect $v^2$ terms, we have
\beqa
\vec{v}\cdot \f{\pa}{\pa t}(\vec{\mathcal{A}}+{\vec{v}}\phi^\pr)\approx -\vec{v}\cdot\(\nabla \phi^\pr+\vec{\mathcal{E}}\)~,~~\vec{v}\tm\f{\pa}{\pa t}(\vec{\mathcal{A}}+{\vec{v}}\phi^\pr)\approx -\vec{v}\tm\(\nabla \phi^\pr+\vec{\mathcal{E}}\).
\eeqa
Changing back from the natural unit, we can obtain the traveling wave type description for Pauli equation
\beqa
&&\[i\hbar\f{\pa}{\pa t}-e \(\phi^\pr+{\vec{v}\cdot\vec{\mathcal{A}}}\)\] \tl{\Psi}\nn\\
&=&-\f{\hbar^2}{2m}\[\nabla-i\f{e}{c} (\vec{\mathcal{A}}+\f{\vec{v}}{c^2}\phi^\pr)\]^2\tl{\Psi}
-\f{e\hbar}{2mc}e\vec{\sigma}\cdot\(\vec{\mathcal{B}}+\f{{\vec{v}}}{c^2}\tm\vec{\mathcal{E}}\)\tl{\Psi}~\nn\\
&-&\f{\hbar^2}{2mc^2}\({\vec{v}}\cdot\[\nabla-ie (\vec{\mathcal{A}}+\f{\vec{v}}{c^2}\phi^\pr)\]\)\f{\pa}{\pa t}\tl{\Psi}\nn\\
&-&\f{ie\hbar^2}{4mc}\left\{\[\vec{v}\cdot \(\nabla \phi^\pr+\vec{\mathcal{E}}\)\]+i\vec{\sigma}\cdot \[\vec{v}\tm\(\nabla \phi^\pr+\vec{\mathcal{E}}\)\]\right\}\tl{\Psi}~,\nn\\
&\approx&-\f{\hbar^2}{2m}\(\nabla-\f{ie\vec{\mathcal{A}}}{c}\)^2\tl{\Psi}
-\f{e\hbar}{2mc}\vec{\sigma}\cdot\vec{\mathcal{B}}\tl{\Psi}
-\f{\hbar^2}{2mc^2}\[{\vec{v}}\cdot\(\nabla-ie \vec{\mathcal{A}}\)\]\f{\pa}{\pa t}\tl{\Psi}\nn\\
&-&\f{ie\hbar^2}{4mc}\left\{\[\vec{v}\cdot \(\nabla \phi^\pr+\vec{\mathcal{E}}\)\]+i\vec{\sigma}\cdot \[\vec{v}\tm\(\nabla \phi^\pr+\vec{\mathcal{E}}\)\]\right\}\tl{\Psi}~.
\eeqa
We can see that the traveling wave description for Pauli equation has many new velocity related new terms. In the $v\ra 0$ limit, this equation can go back to the ordinary Pauli equation. Besides, this equation can not be obtained by naive replacements eq(\ref{Galilean}) in Galilean transformation for $\pa_t$ and $\nabla$.

\section{\label{conclusion}Conclusions}

We derive an equivalent traveling wave form description for Dirac field. In the non-relativistic limit, such form can reduce to inverse-Galilean transformed Schrodinger-type equation. We find that, the resulting two-component Schrodinger-type equation from the reduction of traveling wave form description of Dirac field is different to the naive Galilean transformed Schrodinger equation. Taking into account the interactions of the system to electromagnetic field by adding proper forms of covariant derivative, the traveling wave form description for Pauli equation can be obtained similarly in the non-relativistic limit. This form is also different to the naive Galilean transformed Pauli equation. Such descriptions allow one to choose arbitrary convenient reference frame for quantum system involving spins.

The spin-3/2 Rarita-Schwinger field and spin-2 gravitational field can also have a traveling wave form description as they transform non-trivially with Lorentz transformation. In the Bargmann-Wigner formulation, a field of rest mass $m$ and spin $s\geq 1/2$ can be represented by a completely symmetric multi-spinor $\Psi_{\al_1\cdots\al_{2s}}(x)$ of rank $2s$ that satisfy Dirac-type equations in all its indices.  So, the traveling wave form description for Dirac field can be readily generalized to field with arbitrary spin $s\geq 1/2$, including the spin-3/2 Rarita-Schwinger field and spin-2 gravitational field.

\acknowledgments
This work was supported by the Key Research Project of Henan Education Departm ent for colleges and universities under grant number 21A140025.

\end{document}